\documentclass[10pt, a4paper]{article}

\usepackage[final]{lrec2026} 

\usepackage{times}

\usepackage[T1]{fontenc}

\usepackage[utf8]{inputenc}

\usepackage{microtype}

\usepackage{inconsolata}

\usepackage[utf8]{inputenc} 
\usepackage[T1]{fontenc}    
\usepackage{url}            
\usepackage{booktabs}       
\usepackage{amsfonts}       
\usepackage{microtype}      
\usepackage{lipsum}
\usepackage{fancyhdr}       
\usepackage{multirow}
\usepackage{amsmath}        
\usepackage{comment}
\usepackage{caption}        

\usepackage{fixme}  
\fxsetup{
    status=draft,
    author=,
    layout=inline,
    theme=color
}

\title{Benchmarking Large Pretrained Multilingual Models on Québec French Speech Recognition}

\name{
  Coralie Serrand, Gilles Boulianne, Amira Morsli}
  
\address{
  Computer Research Institute of Montréal, Québec, Canada \\
  coralie.serrand@telecomnancy.net,\{gilles.boulianne,amira.morsli\}@crim.ca\\
}

\abstract{
We evaluate the performance of large pretrained multilingual speech recognition models on a regional variety of French spoken in Québec, Canada, in terms of speed, word error rate and semantic accuracy. To this end we build a benchmark and evaluation pipeline based on the CommissionsQc datasets, a corpus of spontaneous conversations recorded during public inquiries recently held in Québec. Published results for these models on well-known benchmarks such as FLEURS or CommonVoice are not good predictors of the performance we observe on CommissionsQC. Our results should be of interest for practitioners interested in building speech applications for realistic conditions or regional language varieties.
\\ \newline }

\begin{document}
\maketitleabstract

\section{Introduction}

The availability of large multilingual datasets such as MLS (Multilingual LibriSpeech \citep{pratap2020}, VoxPopuli \citep{wang2021}, and Common Voice \citep{ardila2020} have been instrumental in the development of large pretrained speech recognition models. However, these datasets include only the main varieties of English and other languages, so pretrained models may face challenges when processing regional varieties. 

In these datasets, Québec French (QF) represents a negligible portion of the data, given that speakers of this variety constitute approximately 2.3\% of the global French-speaking population \citep{marcoux2024}. We estimate from demographic data available for Common Voice that QF accounts for less than 5\% of the total duration of the French recordings. This low representation may adversely impact the accuracy of speech recognition for QF, and lends support to anecdotal evidence that commonly available models perform less effectively on QF compared to European French.

To quantify this impact, we set out to examine just how well massively multilingual models can handle the Québec variety of French. We built an evaluation pipeline around CommissionsQC~\citep{serrand2025a}, a dataset of QF recently created for training and evaluation of QF speech recognition models, which contains 782 hours of training and 27 hours of development and test data. The recordings and transcripts are derived from two public inquiries that were held recently in Québec, which consist of hearing room interviews of witnesses, leading to a spontaneous, conversation-style speech with some amount of reverberation and background noise.

We measured the performance of large pretrained models (also designated as foundation models) for speech-to-text, in terms of word error rate, semantic similarity, and speed. We also evaluated multimodal models handling both text and speech inputs. All evaluations were done with available pretrained models, without retraining or fine-tuning, to reflect real-life scenarios where task-specific data cannot be used, for privacy concerns or other reasons. Finally, as a reference, we also compared with models trained from scratch on CommissionsQC using 780 hours of QF.

\subsection{Related Work}

The importance of datasets and benchmarks for accented varieties was stressed in  \citep{aksenova2022}. Minority varieties often face reduced ASR performance, for example \citet{koenecke2020} reports that "Word error rate (WER) on a corpus of African American Vernacular English (AAVE) is sometimes observed to be as much as 85\% higher than WER for a corpus of Standard American English".

Some previous studies focused specifically on accented varieties of French.
\citet{maison2023} highlights the challenge of accented speech recognition while showing how to combine multiple accented databases to improve over single accent baselines. \textit{LeBenchmark 2.0} \citep{parcollet2024a} was specifically developed to add spontaneous and accented speech to the initial version of \textit{LeBenchmark}, for pretraining and evaluating SSL models. These two works focus on accented speech but not QF in particular, as the only QF specific dataset they use is CaFE (Canadian French Emotional) \citep{gournay2018}, a one-hour corpus of 
emotional speech among thousands of hours of other French varieties. \citet{zhang2023c} did use a corpus of spontaneous QF from YouTube videos, and found that two variants of Whisper had a zero-shot word error rate much higher than what is reported on benchmarks of European French on a small corpus of 4.5 hours. Recently, \citet{maison2025} created CEREALES, another corpus of QF from public inquiries; they report word error rate results with Whisper small, medium and large model sizes.

We extend these previous works in the following ways: (1) we introduce a new evaluation pipeline for QF speech recognition based on CommissionsQC, a large dataset of spontaneous QF speech, (2) we compare the performance of 24 recent state-of-the-art speech recognition models with diverse architectures on CommissionsQC using the new pipeline, (3) we compare the performance measured for QF using the new pipeline with the performance reported in the literature for European French on FLEURS \cite{conneau2022} and CommonVoice \cite{ardila2020} benchmarks, and (4) we report results on semantic similarity in addition to speed and word error rate.

\section{Experiments}

For back-to-back comparison, all experiments are performed using the same pipeline implementation. All evaluations are done with pretrained models as available, without retraining or fine-tuning, except for two models trained from scratch to establish a baseline for the performance that can be attained when in-domain data is available in sufficient amounts. 

\subsection{Datasets}

We evaluate on the CommissionsQC dataset \citep{serrand2025a}, a corpus of spontaneous, QF created from recordings of two public inquiry hearings. Table~\ref{tab:corpus_stat} summarizes the characteristics of the datasets.
The dataset is split into separate training, development and test subsets without speaker overlap. Results are reported on the development and test subsets of both Bast and Charb parts of the corpus, which are gender balanced.

\begin{table*}[ht]
     \small
     \centering
        \begin{tabular}{llrrrrrr}
        \toprule
                       &  & Speech & N. utt. & N. words &  N. speakers & N. female & Dur. female \\
         Dataset &  Split & dur. (h)         &         &          &              & (\%) & (\%) \\
        \midrule
        \multirow[t]{3}{*}{Bast} & Dev & 4.0 & 999 & 36.2K & 15 & 40.0 & 50.6 \\
         & Test & 8.0 & 2.24K & 74.1K & 19 & 42.1 & 49.0 \\
         & Train & 72.7 & 29.7K & 714K & 46 & 23.9 & 12.6 \\
        \cmidrule{1-8}
        \multirow[t]{3}{*}{Charb} & Dev & 5.0 & 1.53K & 49.5K & 31 & 48.4 & 49.0 \\
         & Test & 10.0 & 3.6K & 102K & 53 & 49.1 & 50.2 \\
         & Train & 709 & 301K & 7.28M & 339 & 19.8 & 21.9 \\
        \bottomrule
        \end{tabular}
     \caption{CommissionsQC datasets statistics.}
    \label{tab:corpus_stat}
\end{table*}

\subsection{Models}

We evaluated models that are easily available publicly. We looked at fully multilingual models as well as variants fine-tuned specifically for French. The list appears in Table~\ref{tab:models}. Cloud models can only be accessed through a cloud API. Foundation models are multilingual models pretrained for speech-to-text conversion, for which weights are available for training or finetuning. 

From scratch models are trained using only the training subsets of Bast and Charb (comprising a total of 782 hours of transcribed speech). 

Multimodal models are more recent multilingual models that are pretrained for multimodal tasks involving speech, text and, in some cases, images. They take an audio input together with a text prompt for instructions. We provided the audio with a prompt asking for a transcription in French.

Except where  noted, all models were used with their original weights and hyperparameters. In the case of multimodal models, the recommended prompts did not usually lead to good performance, so we had to create better prompts, as described in section \ref{sec:multimodal_models}.

\begin{table*}[htb]
\centering
\small
\begin{tabular}{lllll}
\toprule
Model (reference) & Developer & Params & Type & FT \\
\midrule
aws-fr-CA  & AWS & - & cloud & fr-ca \\
azure-speech  & Microsoft & - & cloud & fr-ca \\
google-chirp \citep{zhang2023b} & Google & - & cloud & fr-ca \\
gpt-4o-transcribe  & OpenAI & - & cloud & fr \\
\midrule
canary-1b-flash \citep{zelasko2025} & nvidia & 1B & foundation & none \\
fastconformer\_fr \citep{rekesh2023} & NVIDIA & 115M & foundation & fr \\
faster-whisper-medium  & Systran & 769M & foundation & none \\
lebenchmark-7K-cv \citep{macaire2024a} & Propicto & 315M & foundation & fr \\
mms-1b-all \citep{pratap2023a} & Facebook & 1B & foundation & none \\
mms-1b-fl102 \citep{pratap2023a} & Facebook & 1B & foundation & none \\
mms-1b-l1107 \citep{pratap2023a} & Facebook & 1B & foundation & none \\
w3-large-v3-fr-d16  & Zaion & 1.55B & foundation & fr \\
whisper-base \citep{radford2023} & OpenAI & 74M & foundation & none \\
whisper-large-v2 \citep{radford2023} & OpenAI & 1.55B & foundation & none \\
whisper-large-v3 \citep{radford2023} & OpenAI & 1.55B & foundation & none \\
whisper-large-v3-french  & Zaion & 1.55B & foundation & fr \\
whisper-large-v3-turbo \citep{radford2023} & OpenAI & 1.55B & foundation & none \\
whisper-medium \citep{radford2023} & OpenAI & 769M & foundation & none \\
whisper-small \citep{radford2023} & OpenAI & 244M & foundation & none \\
whisper-tiny \citep{radford2023} & OpenAI & - & foundation & none \\
\midrule
chain\_B+C\_french \citep{han2021} & CRIM & 225M & from scratch & fr-ca \\
espnet\_transformer \citep{watanabe2020} & CRIM & 29M & from scratch & fr-ca \\
\midrule
Phi-4-multimodal-instruct \citep{abouelenin2025} & Microsoft & 5.6B & multimodal & none \\
Qwen2-Audio-7B \citep{chu2024b} & Alibaba Cloud & 8B & multimodal & none \\
gemini-2.0-flash \citep{team2024b} & Google Deepmind & - & multimodal & none \\
seamless-m4t-v2-large \citep{loicbarrault2023} & Meta & 2.3B & multimodal & none \\
\bottomrule
\end{tabular}
\caption{Models evaluated. The FT column indicates if finetuned on European French (fr) or QF (fr-ca).}
\label{tab:models}
\end{table*}

\subsection{Evaluation pipeline}

\begin{figure}[ht]
  \centering
    \includegraphics[width=\linewidth]{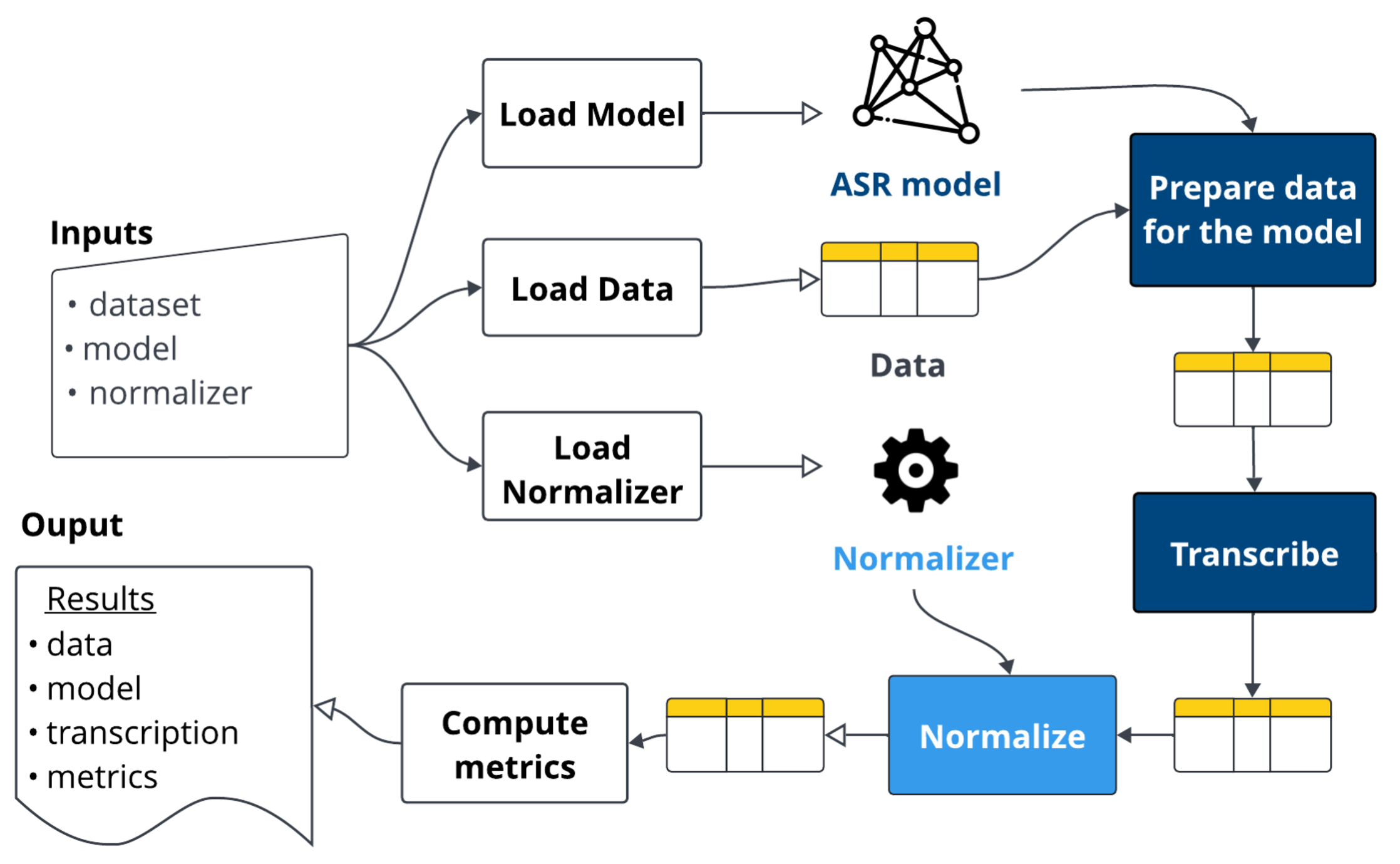}
    \caption{Evaluation pipeline.}
    \label{fig:pipeline}
\end{figure}

To evaluate many models available through multiple implementations and APIs, we created a unified pipeline to handle all model types. It applies the same pre- and post- processing steps in all evaluations, making the results comparable across models and datasets. The data flow is illustrated in Figure~\ref{fig:pipeline}. 

The input to the pipeline is a dataset, a model and a normalizer. Once these are loaded, the data preparation and transcription steps are executed. The resulting transcribed text is normalized to compensate for differences in orthographic variants, as described in section \ref{sec:perf_metrics}, before computing evaluation metrics such as word error rate. The only model-dependent elements are the data preparation and the transcription steps so a new model can be evaluated by coding only these steps and running the pipeline.

\subsection{Performance Metrics} \label{sec:perf_metrics}

In addition to word and character error rates, the pipeline computes RTF (speed) and a Bert F1, a semantic similarity metric.

\paragraph{Word or Character Error Rate} are the most used in speech recognition work. They are computed from the Levenstein edit distance between the reference and transcript text using the jiwer\footnote{\url{https://jitsi.github.io/jiwer/}} library, which also reports the number of substituted, inserted and deleted words (or characters) as well as several other derived measures.

\paragraph{Real-Time Factor (RTF)} is the ratio of processing time to audio duration, measured while processing one utterance at a time i.e., with a batch size equal to 1, corresponding to real-time or streaming processing. The time does not include audio file transfer to cloud but does include transcription file transfer from cloud to client. It can be expressed in \%, so that a 25\% RTF means it took $\frac{1}{4}$ of a second to process 1 second of speech.

\paragraph{Bert F1:} the semantic fidelity of the transcription depends on which words are most affected by recognition errors. To measure how much of the original meaning is captured in the recognition output, we use Bert F1 \citep{zhang2020b}, which is based on the similarity of contextual embeddings of tokens in the reference and test utterances. This metric is well correlated with human judgment of semantic equivalence and is reported here as a value between 0 and 100. For scoring our French text, the embeddings are computed with the multilingual BERT$_{\mathrm{multi}}$ model.

\paragraph{Text normalization:} for WER, CER or Bert F1 evaluation, both the transcript text and the reference text are processed to standardize spelling and punctuation as much as possible, so that two words match regardless of nonsignificant differences.

For the pipeline, text normalization is a parameter with value \textit{basic} or \textit{whisper}. \textit{Whisper} normalization includes lower-casing, removing words within brackets or parentheses, and removing punctuation. \textit{Basic} normalization includes \textit{whisper} normalisation but adds more French specific processing: spaces after apostrophes, splitting of compound words (containing hyphens), separation of digits from their units, and conversion of ordinal and cardinal numbers into their letter format. 

In this paper, results are reported with the \textit{basic} normalization, as it handles correctly acceptable variants in French that should not be considered as errors. 
The impact of either normalization on the WER and Bert F1 is detailed in section \ref{sec:text_norm_impact}.

\section{Results}

Performance evaluation results appear in Table~\ref{tab:aggregated_results}, as aggregated over the four sets of Bast and Charb development and test sets, comprising a total 262K words. More detailed results are provided in Appendix \ref{sec:appendix}. 

\begin{table}[htb]
\centering
\small
\begin{tabular}{lrrrr}
\toprule
       & WER & CER & RTF & Bert \\
 Model & (\%) &  (\%)  &  (\%) &  F1 \\
\midrule
espnet\_transformer & 8.2 & 3.8 & 36 & 95.8 \\
whisper-large-v3-turbo & 8.2 & 4.6 & 6 & 95.4 \\
whisper-large-v3 & 8.4 & 4.9 & 14 & 95.8 \\
w3-large-v3-fr-d16 & 9.2 & 5.0 & 9 & 95.8 \\
whisper-large-v3-french & 10.1 & 5.7 & 14 & 95.6 \\
aws-fr-CA & 10.3 & 5.0 & 84 & 94.7 \\
faster-whisper-medium & 10.7 & 6.0 & 11 & 94.2 \\
whisper-medium & 12.8 & 7.8 & 9 & 92.3 \\
azure-speech & 13.4 & 6.0 & 46 & 90.7 \\
Phi-4-multimod-instruct & 13.4 & 6.9 & 31 & 93.1 \\
gemini-2.0-flash & 13.7 & 7.2 & 20 & 92.1 \\
gpt-4o-transcribe & 14.2 & 10.8 & 17 & 93.8 \\
whisper-large-v2 & 14.7 & 9.7 & 14 & 92.6 \\
chain\_B+C\_french & 15.3 & 6.5 & 22 & 94.4 \\
google-chirp & 15.4 & 8.3 & 28 & 91.4 \\
fastconformer\_fr & 20.6 & 13.3 & 3 & 90.3 \\
lebenchmark-7K-cv & 24.0 & 10.3 & 1 & 88.8 \\
canary-1b-flash & 26.0 & 17.9 & 4 & 87.7 \\
mms-1b-all & 28.7 & 12.7 & 3 & 85.1 \\
seamless-m4t-v2-large & 31.0 & 24.3 & 16 & 88.5 \\
whisper-small & 33.0 & 22.0 & 10 & 88.5 \\
mms-1b-l1107 & 39.5 & 15.8 & 3 & 82.1 \\
mms-1b-fl102 & 42.8 & 18.9 & 3 & 75.3 \\
Qwen2-Audio-7B & 43.2 & 32.5 & 14 & 82.2 \\
whisper-base & 53.7 & 34.7 & 3 & 84.3 \\
whisper-tiny & 74.2 & 44.1 & 3 & 78.8 \\
\bottomrule
\end{tabular}
\caption{Results aggregated over Bast and Charb dev and test sets. WER, CER and RTF in \%.}
\label{tab:aggregated_results}
\end{table}

The best performing model, in terms of error rate as well as Bert F1 score, is \texttt{espnet\_transformer}, a model trained from scratch on the Bast and Charb training sets, which sets a reference for the case where 780 hours of training data are available for a regional variety. Closely following with almost the same performance are the larger \texttt{whisper} versions, \texttt{whisper-large-v3-turbo} being the fastest. We observe that French fine-tuned versions \texttt{whisper-large-v3-french} and \texttt{w3-large-v3-fr-d16} actually perform worse than their original multilingual versions.

Bert F1 has a narrow range but is generally consistent with WER. Models that obtain 10\% or less WER obtain over 94 in Bert F1. One notable exception is \texttt{azure-speech} which has about 13\% WER but a Bert F1 of around 90, which is more typical of models with more than 20\% WER, indicating that its errors are more semantically hurtful than other models in the same word error rate range.

\subsection{Insertions and deletions}

Whisper models tend to insert words or repetitive sequences of words that are not present in the audio, leading to more insertion errors, especially smaller versions. For example, if we look at the deletions/insertions ratio (DIR) on Bast dev, we find that \texttt{chain\_B+C\_french}, a non-transformer chain-TDNN model has a typical DIR of 2.2, while \texttt{whisper-large-v3} DIR decreases to 2.05, \texttt{whisper-medium} to 1.1 and \texttt{whisper-small} to 0.50, indicating increased levels of insertion errors.

\texttt{gpt-4o-transcribe} is a special case: its overall 14.2\% WER is respectable, however its DIR is 20.2, an extremely large value implying lots of deletions. Closer examination shows it frequently skips beginning or ending utterances in a speaker turn, leading to abnormally high deletion rates, while having almost perfect results outside the skipped parts, including very few insertions. If we naively cancel the effect of these skips, by setting the number of deletions equal to insertions, we get a rough estimate of about 3\% WER, which would be the lowest error rate by far. This weird behavior has been previously reported on OpenAI developer community forums and maybe due to post-training model alignment, but the cause remains unknown.

\subsection{Speed}

\begin{figure*}[htb]
  \centering
    \includegraphics[width=0.7\linewidth]{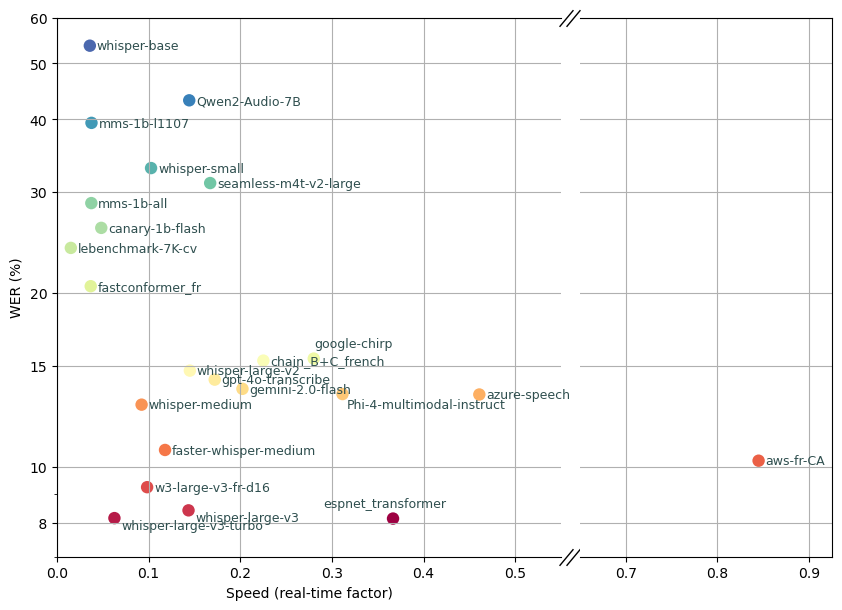}
    \caption{Word error rate as a function of RTF for the tested models.}
    \label{fig:rtf_vs_wer}
\end{figure*}

Speed measurements appear in the fourth column of Table~\ref{tab:aggregated_results}, as RTF \% (real-time factor in \%). Non-cloud models were run locally on RTX2080Ti GPUs, except  \texttt{Phi-4}, \texttt{Qwen2}, and \texttt{espnet\_transformer} which ran on an A40 due to memory constraints. All models were running faster than real-time (RTF less than 100\%). The relationship between speed and error rate is easier to visualize in Figure~\ref{fig:rtf_vs_wer}. We observe a cluster, around 0.2 to 0.3 RTF and 15\% WER, which includes models from a large variety of types and architectures, which could be said to represent typical speech recognition performance on QF at the time when these results were obtained.

\subsection{Gender bias}

Globally we observe only a slight gender imbalance in the results, in favor of female, with \%WER being 3\% lower (relative) for female speakers. The best performer, \texttt{whisper-large-v3-turbo}, yields 8.4\% WER for male and 8.0\% WER for female (a 5\% relative difference). Only a minority of models are better for male than female speakers, with no clear trend according to model type. Table \ref{tab:gender_results} in Appendix \ref{sec:appendix} gives a more detailed account of gender-based results, aggregated over the two datasets.

\subsection{Cloud services}

Models identified as "cloud" in Table~\ref{tab:models} are available through online transcription services. In all cases we used the "fr-CA" version. 
We used standard APIs rather than real-time APIs so in some cases processing time includes waiting for available processing power. WER ranges from 10\% for \texttt{aws-fr-CA} to 15\% WER for \texttt{google-chirp}. The fastest cloud model is \texttt{gpt-4o-transcribe} at 17\% RTF, which is 3 times slower than open-source \texttt{whisper-large-v3-turbo}. Notably, \texttt{aws-fr-CA} and \texttt{azure-speech} are the slowest of all models\footnote{AWS has a "fast transcription service" for some languages, but it is not available for the "fr-CA" language at this time.}. 

\subsection{Multimodal models} \label{sec:multimodal_models}

We tested models that can handle audio and text as input, listed as multimodal in the last rows of Table~\ref{tab:models}. Of these four, \texttt{Phi-4} and \texttt{gemini-2.0} yield less than 14\% WER, on a par with speech only models. However, \texttt{seamless-m4t} and \texttt{Qwen2-Audio}, although being the largest of all evaluated models, in number of parameters, perform poorly with more than 30\% WER.

We observed that multimodal model transcription output is sensitive to details in the prompted instruction. For each model we experimented with a least 3 different prompts and selected the one which obtained the best performance on Bast development set. The recommended prompts yielded poor performance for this QF transcription task. For \texttt{Phi-4-multimodal-instruct}, for example, the word error rate was reduced from 25.1\% with the original prompt, to 15.3\% with the best prompt. Table \ref{tab:prompts} in Appendix \ref{sec:appendix} gives the detailed results from this prompt experiment. For \texttt{Gemini-2.0} and \texttt{Phi-4} a french prompt worked best, but an english one was better for \texttt{Qwen2-Audio}. For \texttt{seamless-m4t-v2} we used the "SpeechToText" version of the model, so no text prompt was needed.

\subsection{FLEURS and CommonVoice}

\begin{figure}[htb]
  \centering
    \includegraphics[width=\linewidth]{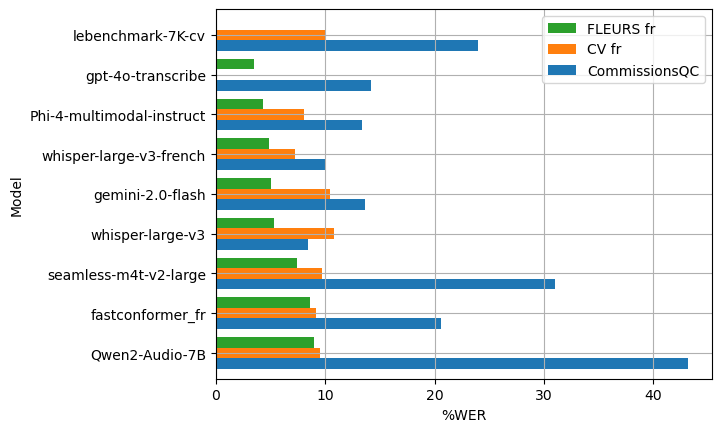}
    \caption{Performance on french FLEURS, french CommonVoice and CommissionsQC (averaged over Bast and Charb development and test sets), sorted in order of increasing error rate on FLEURS.}
    \label{fig:cq_vs_cv_fleurs}
\end{figure}

Performance on the French data subsets of standard benchmarks FLEURS and CommonVoice (CV) is available for some of the models evaluated here, in model cards or in the literature  \cite{abouelenin2025,pratap2023a,parcollet2024a}.
Figure~\ref{fig:cq_vs_cv_fleurs} compares published results on these benchmarks with our results on CommissionsQC, for nine models. 

The first observation is that word error rates on CommissionsQC are much higher than on the benchmarks. This is somewhat expected because models tend to be optimized for benchmarks such as FLEURS and CommonVoice, which are clean, read speech corpora, while CommissionsQC is spontaneous speech in court room conditions. 

We also find that the benchmarks do not predict at all how the models will rank on CommissionQC. The best performing model on FLEURS, \texttt{gpt-4o}, ranks 5th out of 9 on CommissionsQC, while \texttt{whisper-large-v3}, which ranks 1st on CommissionsQC, ranks 6th out of 9 on FLEURS. More strikingly, the worst performing model on CV is \texttt{whisper-large-v3}, which is the best performing model on CommissionsQC.

\subsection{Impact of text normalization} \label{sec:text_norm_impact}

Text normalization is applied to both reference text and recognizer output, as mentioned in section \ref{sec:perf_metrics}, and thus affects both \% WER and Bert F1 metrics. To judge its impact on results reported here and elsewhere, we show in Table \ref{tab:basic_vs_whisper} a comparison of Bert-F1 and \% WER metrics of the output of \texttt{whisper-large-v3} for the two normalizations, aggregated over Bast and Charb development and test sets. 
There is about 1\% absolute difference for word error rate, and a smaller effect on Bert F1. Since published results probably used \textit{whisper} normalization, the word error rate prediction gap with CommissionsQC shown in Figure~\ref{fig:cq_vs_cv_fleurs} is probably underestimated (i.e. it would appear even larger if we had used \textit{whisper} normalization).

\begin{table}[ht]
\centering
\begin{tabular}{lrr}
\toprule
Normalization  & \%WER & BERT F1 \\
\midrule
\textit{basic} & 8.17 & 95.88 \\
\textit{whisper} & 9.39 & 95.15 \\
\bottomrule
\end{tabular}
\caption{Effect of text normalization on evaluation metrics for \texttt{whisper-large-v3.}}
\label{tab:basic_vs_whisper}
\end{table}

\section{Conclusion} 

We evaluated 24 pretrained multilingual speech recognition models on CommissionsQC, a Québec French benchmark of public inquiries, in terms of speed, word error rate and semantic accuracy. Our results should be of interest for research and applications in Québec French, in spontaneous and realistic ambient settings, and for practitioners that are looking for guidance on model selection to meet specific computational, storage and performance constraints. For spontaneous Québec French in hearing room conditions, the best current publicly available model yields 8\% WER in 0.06 times real-time, while typical models cluster at around 14\% WER and 0.2 RTF. In particular, we found that model performance reported for French on standard benchmarks like FLEURS or CommonVoice is not a good predictor of actual performance for the QF regional variety, style and recording conditions of our benchmark.
Our next work should explore strategies for fine-tuning on CommissionsQC data in order to improve over the baseline performances established here.

\section{Limitations}
This study has several limitations related to both the nature of the dataset and the evaluation setup. A key constraint is the difficulty of isolating specific factors that influence model performance. The CommissionsQC corpus combines multiple challenging dimensions—including a regional variety of French, spontaneous speech, and realistic courtroom conditions such as background noise, reverberation, and speaker overlap. As a result, it is not possible to attribute recognition errors to any single cause with certainty.

Although CommissionsQC offers a representative sample of Québec French in a formal institutional setting, it remains limited in scope. Informal registers, youth or rural speech varieties, telephone conversations, and everyday dialogue are not represented, which restricts the generalizability of the findings.

There is also a small risk of data contamination. While we have verified that CommissionsQC recordings are no longer publicly accessible, we cannot entirely rule out the possibility that some data may have been indirectly seen during pretraining of certain models. That said, this appears unlikely, especially for the audio recordings, which were removed from the web several years ago, but the lack of transparency in large-scale model training leaves some uncertainty.

Multimodal models introduce another challenge, as their performance proved highly sensitive to prompt formulation. Choosing prompts based on development data introduces a degree of subjectivity and may limit reproducibility.

Finally, some models were accessed via proprietary cloud APIs (e.g., GPT-4o, AWS, Google Speech), which do not guarantee version stability. This means that results obtained at evaluation time may not be fully replicable in the future, even with identical API calls.

\section{Ethical considerations}

This work raises ethical concerns primarily related to the nature of the audio data and the absence of explicit speaker consent.

Although the recordings in CommissionsQC were originally captured during public hearings and are part of the public record, the individuals recorded did not provide informed consent for their voices to be reused in research or for model training. This lack of explicit consent limits the permissible scope of data reuse and requires particular caution.

Furthermore, voice is a biometric identifier and must be treated as personal data. Its misuse can lead to serious harms, including unauthorized voice reproduction, circumvention of voice authentication systems, and the generation of deepfake audio or synthetic impersonations.

To mitigate these risks, the CommissionsQC dataset is not currently released for open access. Its potential public distribution is under review and, if approved, would be restricted to research use under a license that includes: manual validation of access requests and an explicit prohibition against using the dataset for training or generating synthetic voices or vocal impersonations.

These measures aim to prevent misuse of the data while supporting legitimate research on regional speech recognition.

\section*{Acknowledgements}

We acknowledge the support of the Natural Sciences and Engineering Research Council of Canada (NSERC) for this work 
and would also like to thank Ministry of Economy and Innovation (MEI) of the Government of Québec for its continued support. 

\vfill

\nocite{*}
\section{Bibliographical References}

\bibliographystyle{lrec2026-natbib}
\bibliography{fq_trunc5}

\appendix
\onecolumn

\section{Additional tables} \label{sec:appendix}

\begin{table*}[htb]
\centering
\begin{tabular}{lrrrrrrrr}
\toprule
 & \multicolumn{4}{c}{\%WER} & \multicolumn{4}{c}{RTF} \\ \cmidrule(l){2-5} \cmidrule(l){6-9}
Model \hfill Dataset  & Bdev & Btest & Cdev & Ctest & Bdev & Btest & Cdev & Ctest \\
\midrule
whisper-large-v3-turbo & 6.30 & 7.07 & 7.66 & 9.86 & 0.05 & 0.06 & 0.06 & 0.08 \\
whisper-large-v3 & 7.54 & 7.32 & 8.15 & 9.66 & 0.08 & 0.18 & 0.09 & 0.21 \\
espnet\_transformer & 7.87 & 7.79 & 7.97 & 8.59 & 0.38 & 0.37 & 0.35 & 0.35 \\
w3-large-v3-fr-d16 & 7.93 & 8.29 & 8.97 & 10.51 & 0.06 & 0.12 & 0.08 & 0.14 \\
aws-fr-CA & 8.52 & 8.82 & 10.73 & 11.70 & 0.92 & 0.73 & 0.92 & 0.81 \\
whisper-large-v3-french & 8.73 & 10.11 & 8.95 & 11.07 & 0.08 & 0.19 & 0.09 & 0.23 \\
faster-whisper-medium & 8.73 & 9.56 & 10.79 & 12.21 & 0.11 & 0.11 & 0.12 & 0.13 \\
Phi-4-multimodal-instruct & 9.81 & 15.34 & 16.07 & 11.94 & 0.30 & 0.28 & 0.33 & 0.33 \\
whisper-large-v2 & 9.99 & 10.27 & 16.61 & 18.64 & 0.08 & 0.19 & 0.10 & 0.21 \\
whisper-medium & 11.41 & 10.20 & 13.85 & 14.75 & 0.04 & 0.13 & 0.05 & 0.15 \\
azure-speech & 11.89 & 12.37 & 12.69 & 14.93 & 0.45 & 0.45 & 0.47 & 0.47 \\
gpt-4o-transcribe & 11.95 & 13.46 & 12.54 & 16.28 & 0.15 & 0.16 & 0.17 & 0.20 \\
chain\_B+C\_french & 12.21 & 13.63 & 15.88 & 17.31 & 0.21 & 0.21 & 0.24 & 0.24 \\
gemini-2.0-flash & 12.26 & 12.60 & 13.19 & 15.17 & 0.14 & 0.21 & 0.21 & 0.25 \\
google-chirp & 12.97 & 14.39 & 14.62 & 17.42 & 0.26 & 0.27 & 0.29 & 0.30 \\
fastconformer\_fr & 17.09 & 18.09 & 21.30 & 23.26 & 0.03 & 0.03 & 0.04 & 0.05 \\
canary-1b-flash & 20.02 & 22.44 & 26.36 & 30.42 & 0.05 & 0.05 & 0.05 & 0.05 \\
mms-1b-all & 23.60 & 24.38 & 28.35 & 33.69 & 0.04 & 0.04 & 0.04 & 0.04 \\
whisper-small & 28.22 & 36.83 & 28.40 & 34.02 & 0.03 & 0.08 & 0.03 & 0.28 \\
seamless-m4t-v2-large & 30.18 & 29.33 & 34.07 & 31.12 & 0.15 & 0.16 & 0.17 & 0.19 \\
mms-1b-l1107 & 33.67 & 34.94 & nan & 44.77 & 0.04 & 0.04 & nan & 0.04 \\
mms-1b-fl102 & 35.34 & 38.34 & 42.86 & 48.70 & 0.04 & 0.04 & 0.04 & 0.04 \\
Qwen2-Audio-7B & 41.80 & 45.24 & 37.16 & 45.06 & 0.14 & 0.14 & 0.15 & 0.15 \\
whisper-base & 42.12 & 47.93 & 56.29 & 60.64 & 0.02 & 0.05 & 0.02 & 0.06 \\
whisper-tiny & 58.53 & 74.01 & 71.14 & 81.47 & 0.02 & 0.04 & 0.02 & 0.05 \\
\bottomrule
\end{tabular}
\caption{Detailed results for dev and test subsets, ordered by ascending \%WER on Bdev.}
\label{tab:detailed_results}
\end{table*}

\begin{table*}[htb]
\centering
\begin{tabular}{lrrrrrr}
\toprule
      & \multicolumn{3}{c}{\%WER} & \multicolumn{3}{c}{\%CER} \\ \cmidrule(l){2-4} \cmidrule(l){5-7}
Model & avg   & male  & female & avg & male & female \\
\midrule
espnet\_transformer & 8.2 & 8.5 & 7.8 & 3.8 & 3.9 & 3.7 \\
whisper-large-v3-turbo & 8.2 & 8.4 & 8.0 & 4.6 & 4.7 & 4.5 \\
whisper-large-v3 & 8.4 & 8.4 & 8.5 & 4.9 & 4.9 & 5.0 \\
w3-large-v3-fr-d16 & 9.2 & 9.1 & 9.4 & 5.0 & 4.9 & 5.2 \\
whisper-large-v3-french & 10.1 & 10.2 & 10.0 & 5.7 & 5.7 & 5.7 \\
aws-fr-CA & 10.3 & 10.2 & 10.3 & 5.0 & 5.0 & 5.0 \\
faster-whisper-medium & 10.7 & 10.9 & 10.5 & 6.0 & 6.1 & 5.9 \\
whisper-medium & 12.8 & 13.5 & 12.1 & 7.8 & 8.4 & 7.2 \\
azure-speech & 13.4 & 13.7 & 13.0 & 6.0 & 6.2 & 5.8 \\
Phi-4-multimodal-instruct & 13.4 & 14.7 & 12.0 & 6.9 & 7.9 & 5.8 \\
gemini-2.0-flash & 13.7 & 13.9 & 13.5 & 7.2 & 7.3 & 7.1 \\
gpt-4o-transcribe & 14.2 & 15.2 & 13.1 & 10.8 & 11.7 & 9.7 \\
whisper-large-v2 & 14.7 & 15.9 & 13.3 & 9.7 & 10.3 & 8.9 \\
chain\_B+C\_french & 15.3 & 15.5 & 15.1 & 6.5 & 6.6 & 6.5 \\
google-chirp & 15.4 & 16.0 & 14.8 & 8.3 & 8.6 & 8.0 \\
fastconformer\_fr & 20.6 & 20.1 & 21.1 & 13.3 & 12.8 & 13.7 \\
canary-1b-flash & 26.0 & 27.1 & 24.8 & 17.9 & 19.1 & 16.6 \\
mms-1b-all & 28.7 & 28.0 & 29.3 & 12.7 & 12.4 & 13.0 \\
seamless-m4t-v2-large & 31.0 & 30.9 & 31.0 & 24.3 & 24.5 & 24.0 \\
whisper-small & 33.0 & 31.7 & 34.1 & 22.0 & 21.2 & 22.8 \\
mms-1b-l1107 & 39.5 & 39.9 & 39.0 & 15.8 & 15.9 & 15.6 \\
mms-1b-fl102 & 42.8 & 42.5 & 43.3 & 18.9 & 18.6 & 19.3 \\
Qwen2-Audio-7B & 43.2 & 42.6 & 43.8 & 32.5 & 31.9 & 33.3 \\
whisper-base & 53.7 & 54.7 & 52.8 & 34.7 & 35.0 & 34.6 \\
whisper-tiny & 74.2 & 77.3 & 70.8 & 44.1 & 45.9 & 42.2 \\
\bottomrule
\end{tabular}
\caption{Results by gender, aggregated over Bast and Charb dev and test sets.}
\label{tab:gender_results}
\end{table*}

\begin{table*}[htb]
\centering
\begin{tabular}{lp{0.8\linewidth}r}
\toprule
 & Prompt & \%WER \\
\# &  &  \\
\midrule
1 & À partir de l'audio ci-joint, générez en français une transcription textuelle complète du contenu parlé. & 15.3 \\
2 & Transcrire le clip audio en texte français. & 16.1 \\
3 & À partir de l'audio ci-joint, générer en français une transcription textuelle complète du contenu parlé. & 16.5 \\
4 & Faire une transcription textuelle complète en français du contenu parlé du clip audio. & 17.4 \\
5 & Transcribe the french audio clip into text. Transcribe only what is present in the audio clip. Always transcribe in French.  Result: & 18.1 \\
6 & Transcrire tout le clip audio en texte français. Ne pas ajouter de texte supplémentaire. & 18.5 \\
7 & Transcribe the audio clip into text. & 19.1 \\
8 & Transcribe the french audio clip into text. Always transcribe the audio clip in French. Do not add any other text. Do not use any other language. Result: & 24.0 \\
9 & Transcribe the french audio clip into French text. Transcribe all and nothing but what is present in the audio clip. Result: & 25.1 \\
\bottomrule
\end{tabular}
\caption{Tested prompts for Phi-4-multimodal-instruct and resulting \%WER on Bast test set.}
\label{tab:prompts}
\end{table*}

\end{document}